\documentclass{emulateapj}

\shorttitle{M31's Metal-Poor Halo}
\shortauthors{Kalirai et al.}

\begin{document}


\title{The Metal-Poor Halo of the Andromeda Spiral Galaxy (M31)}



\author{
Jasonjot S.\ Kalirai\altaffilmark{4,5},
Karoline M.\ Gilbert\altaffilmark{4},
Puragra Guhathakurta\altaffilmark{4},
Steven~R.~Majewski\altaffilmark{6}, \\
James~C.~Ostheimer\altaffilmark{6},
R.~Michael~Rich\altaffilmark{7}
Michael~C.~Cooper\altaffilmark{8},
David~B.~Reitzel\altaffilmark{7}, \\ and
Richard~J.~Patterson\altaffilmark{6}
}


\altaffiltext{1}{Data presented herein were obtained at the W.\ M.\ Keck
Observatory, which is operated as a scientific partnership among the
California Institute of Technology, the University of California, and the
National Aeronautics and Space Administration.  The Observatory was made
possible by the generous financial support of the W.\ M.\ Keck Foundation.}
\altaffiltext{2}{Based on observations obtained with MegaPrime/MegaCam, a
joint project of CFHT and CEA/DAPNIA, at the Canada-France-Hawaii Telescope
(CFHT) which is operated by the National Research Council (NRC) of Canada,
the Institut National des Science de l'Univers of the Centre National de la
Recherche Scientifique (CNRS) of France, and the University of Hawaii.}
\altaffiltext{3}{Based on observations obtained with the Kitt Peak National 
Observatory.  Kitt Peak National Observatory of the National 
Optical Astronomy Observatories is Operated by the Association of Universities 
for Research in Astronomy, Inc., under cooperative agreement with the National 
Science Foundation.}
\altaffiltext{4}{University of California Observatories/Lick Observatory, 
University of California at Santa Cruz, 1156 High Street, Santa Cruz, 
California 95064, USA; jkalirai@ucolick.org}
\altaffiltext{5}{Hubble Fellow}
\altaffiltext{6}{Department of Astronomy, University of Virginia, P.\ O.\ Box
3818, Charlottesville, Virginia 22903, USA} 
\altaffiltext{7}{Department of Astronomy, University of California at Los Angeles, 
Box 951547, Knudsen Hall, Los Angeles, California 90095, USA}
\altaffiltext{8}{Astronomy Department, 601 Campbell Hall, University of
California at Berkeley, Berkeley, California 94720, USA}


\begin{abstract}

We present spectroscopic observations of red giant branch (RGB) stars over a 
large expanse in the halo of the Andromeda spiral galaxy (M31), acquired with the DEIMOS 
instrument on the Keck~II 10-m telescope.  Using a combination of five 
photometric/spectroscopic diagnostics --- (1)~radial velocity, (2)~intermediate-width 
$DDO51$ photometry, (3)~Na\,{\sc i} equivalent width (surface gravity sensitive), 
(4)~position in the color-magnitude diagram, and (5)~comparison between photometric 
and spectroscopic [Fe/H] estimates --- we isolate over 250 bona fide M31 bulge 
and halo RGB stars located in twelve fields ranging from $R$ = 12--165~kpc from 
the center of M31 (47 of these stars are halo members with $R >$ 60~kpc).  We derive 
the metallicity distribution function of M31 RGB stars in each of these fields by 
comparing the stellar location in the ($I$, $V-I$) color-magnitude diagram to a 
finely spaced grid of theoretical isochrones.  The mean of the resulting M31 
spheroid (bulge and halo) metallicity distribution is found to be 
systematically more metal-poor with increasing radius, shifting from 
$\langle$[Fe/H]$\rangle$ = ${\rm-}$0.47$\pm$0.03 ($\sigma$ = 0.39) at $R <$ 20~kpc 
to $\langle$[Fe/H]$\rangle$ = ${\rm-}$0.94$\pm$0.06 ($\sigma$ = 0.60) at 
$R \sim$ 30~kpc to $\langle$[Fe/H]$\rangle$ = ${\rm-}$1.26$\pm$0.10 
($\sigma$ = 0.72) at $R >$ 60~kpc, assuming [$\alpha$/Fe] = 0.0.  These results 
indicate the presence of a metal-poor RGB population at large radial
distances out to at least $R=160$~kpc, thereby supporting our recent discovery of a
stellar {\it halo\/} in M31: its halo and bulge (defined as the structural
components with $R^{-2}$ power law and de~Vaucouleurs $R^{1/4}$ law surface
brightness profiles, respectively) are shown to have distinct metallicity
distributions.  If we assume an $\alpha$-enhancement of [$\alpha$/Fe] = +0.3 for 
M31's halo, we derive $\langle$[Fe/H]$\rangle$ = ${\rm-}$1.5$\pm$0.1 
($\sigma$ = 0.7).  Therefore, the mean metallicity and metallicity spread of 
this newly found remote M31 RGB population are similar to those of the Milky Way 
halo.

\end{abstract}





\section{Introduction} \label{intro}

Large galaxies such as the Milky Way and Andromeda (M31) are believed to have been 
assembled hierarchically \citep{searle}.  The growth of such galaxies 
is powered by the continual accretion of smaller dwarf galaxies that are tidally 
destroyed as they fall into the larger potential \citep{zentner}.  
Numerical simulations suggest that the most massive merging events occur 
early on ($t_{\rm look  back} \gtrsim$ 8 Gyrs) and form the inner halo 
($R \lesssim$ 20~kpc), whereas the recently accreted dwarfs form structure in 
the halo \citep{bullock2005}.  
This formation scenario leads to several predictions that 
can be observationally tested.  For example, accretion of dwarf galaxies should 
naturally produce an extended stellar halo in massive hosts.  Furthermore, 
the recent infall and subsequent tidal disruption of dwarf satellites should 
produce a large amount of stellar substructure (i.e., tidal streams) still 
existing within galaxy halos.  Finally, since the most massive merging events 
that formed the inner parts of hosts like the Milky Way and M31 are also the most 
metal-rich \citep{font,robertson,renda1,brook}, this formation scenario suggests 
that the inner parts of massive galaxies should be chemically different from their 
halos.

\begin{figure*}
\begin{center}
\leavevmode
\includegraphics[width=14cm]{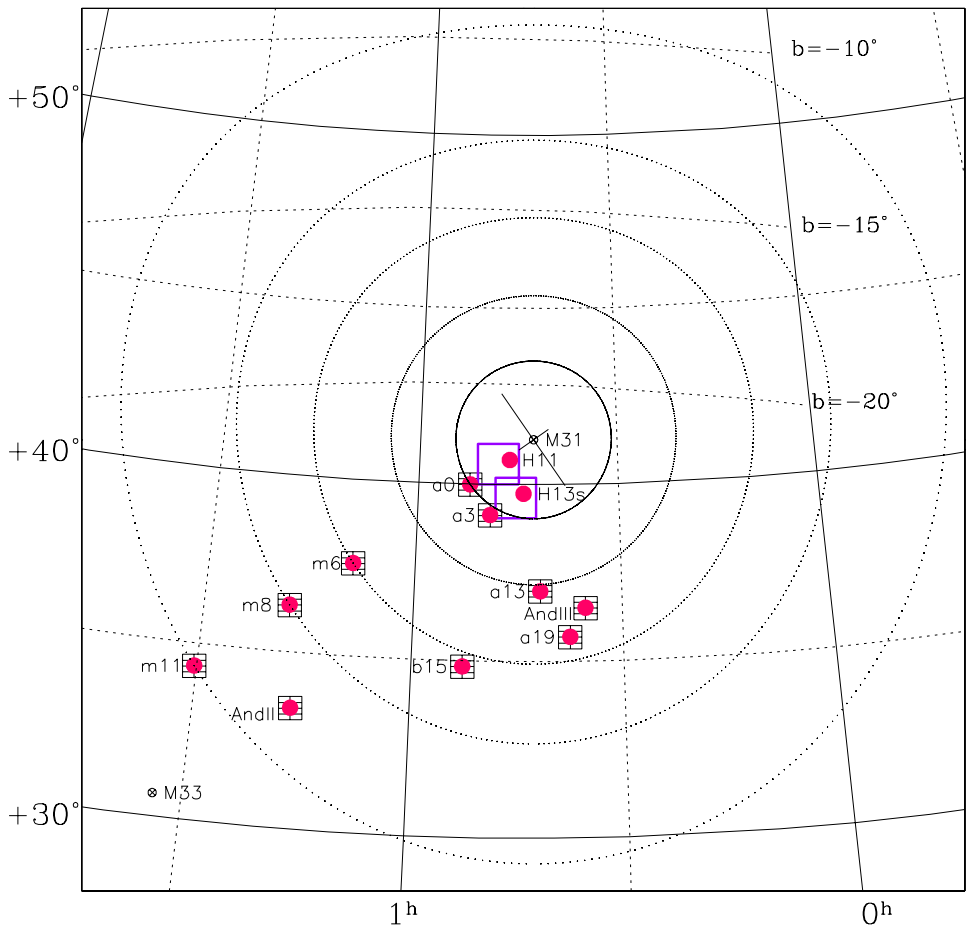}
\end{center}
\caption{Location of the fields used in this study relative to M31 (center).  
The size of M31's visible disk, and orientation, is illustrated with two solid 
lines in the center.  Lines of constant longitude and latitude are also shown.
The smaller gridded squares represent schematically the field centers of the 
KPNO fields for which we have $DDO51$ photometry and the two larger squares 
indicate the approximate positions of the CFHT fields without $DDO51$ photometry.  
The filled circles show the approximate locations of the twelve fields for 
which we have Keck/DEIMOS spectroscopic observations.}
\label{fig:fields}
\end{figure*}

Recently, a large number of ground and space based observations have 
targeted both the Milky Way and M31 to try to confirm these 
predictions directly.  In this respect, M31 is often a better choice than our own 
Galaxy since we have a global, external view of it.  The wealth of data 
collected to date has indeed confirmed certain theoretical predictions, 
while, at the same time, it has raised some new puzzles.  For example, wide-field 
star-count maps of both galaxies have revealed abundant substructure in 
the halos of these systems, as predicted by simulations.  In the Milky Way, 
the Sagittarius stream \citep{ibata94,majewski2}, the Magellanic stream 
\citep{mathewson} and the Monoceros stream \citep{yanny} present clear 
evidence that merging processes are 
prominent in massive galaxies and are still occuring today.  Similarly, 
star-count maps of M31 \citep*[e.g.,][]{ferguson02,ibata05} have shown that its 
spheroid is very imhomogenous with many prominent density enhancements 
\citep*[e.g., the giant southern stream ---][]{ibata01}.

The comparison of the Milky Way and M31 halo has also revealed several 
unanticipated results and notable differences between the two galaxies.  Despite 
their similar overall size, the stellar density of M31's spheroid appears to be 10$\times$ 
higher than that at a comparable location in the Milky Way \citep{reit98}.  
\cite{mould} surprisingly found that the ``halo'' of M31 at $R \sim$ 7~kpc is 
also 10$\times$ more metal rich than the Milky Way's.  This was also measured by 
\cite{durrell94} and by \cite{rich96} over a larger radial distance, from $R$ = 
5.3--19.4~kpc.  Further studies 
of the metallicity distribution function (MDF) in M31 not only confirmed this result, but 
also showed that there is no evidence for an abundance gradient out to 30~kpc 
(Durrell, Harris, \& Pritchet 2001, 2004, Bellazzini et al. 2003), a result at 
odds with model predictions.  Differences 
between the Milky Way and M31 are also seen in the distribution of the ages of stars at 
large radii.  The canonical picture of an old, metal-poor population simply does not appear to
describe M31's ``halo''.  \cite{brown} used the Advanced Camera for Surveys
(ACS) on the {\it Hubble Space Telescope\/} ({\it HST\/}) to target a
minor-axis field at $R\sim10$~kpc in M31 and produced a superb
color-magnitude diagram (CMD) extending well below the main-sequence turnoff.
Their data suggest a broad distribution of stellar ages in M31's ``halo'': over 
half the population of stars in this field have ages $\lesssim11$~Gyr old.  By mass, 
30\% of the stars are found to be 6--8 Gyr old.  Rich et al. (2006, in preparation) 
have now obtained spectroscopy of stars in this {\it HST\/} field using the DEIMOS 
spectrograph on the Keck~II telescope.  We use these data below to confirm 
that the RGB population in the Brown et al. M31 ``halo'' field is much more 
metal rich than RGB stars in the Milky Way halo.  

Until recently, there appeared to be another striking structural difference
between the Milky Way and M31 spheroids.  Whereas the Milky Way shows a 
de Vaucouleurs $R^{1/4}$ \citep{deV58} surface brightness profile in the 
inner regions of the Galaxy and a power law $R^{-2}$ projected profile for 
the halo in the outer parts, \cite{pri94} found that the entire spheroid 
of M31 could be fit by a single $R^{1/4}$ profile.  Their star-count measurements 
extended out to $R \sim$ 20 kpc along the south-east minor axis of M31, and demonstrated
that the surface brighness profile of M31 falls off very steeply with
increasing radius near the limit of their survey.  Similarly, \cite{durrell04} 
found that an $R^{1/4}$ law is consistent with M31's surface brightness profile out 
to 30~kpc.  

\cite{ostheimer} carried out a survey that was a significant improvement over 
previous work in terms of both spatial coverage ($R \sim$ 10--165 kpc), 
photometric depth (1.5 mag arcsec$^{\rm -2}$ fainter), and signal-to-background.  
His M31 surface brightness profile showed the first sign of a flattening (e.g., 
deviation of the slope from a de Vaucouleurs profile) in the outermost few 
bins ($>$80~kpc).  By following up the \cite{ostheimer} imaging observations 
with multiobject Keck/DEIMOS spectroscopy, we have now been able to verify 
M31 red giant branch (RGB) member stars in each of the Ostheimer fields out to
$R \sim$ 160 kpc.  The spectroscopic data unequivocally show a {\it break\/} in 
the surface brightness profile of M31.  In \cite{guh06a}, we present 
the first detection of an $R^{-2}$ {\it halo\/} of stars in M31 extending 
out to a projected radius $>$160 kpc.  Based solely on photometric data out 
to $R \sim$~55 kpc, \cite{irwin} also find a break in the surface brightness 
profile of M31 (although they claim that the outer component does not 
resemble a population II {\it halo\/}).

Following up on this discovery, this paper presents the first detection 
of a metallicity gradient in M31\footnote{Although \cite{ostheimer} did not claim to 
have seen an obvious/significant metallicity gradient in M31, he did provide 
evidence for an increase in the fraction of metal-poor stars in his outermost 
annuli.}.  Furthermore, we show that the crossover between the metal-rich and 
metal-poor components in our sample occurs at a minor-axis distance of 
$\sim$30 kpc, in excellent agreement with the transition 
radius that separates the newly discovered halo of M31 from the inner de 
Vaucouleurs ($R^{1/4}$) {\it bulge\/}\footnote{Throughout the rest of this paper, we use 
the terms ``bulge'' and ``halo'' to refer to M31's $R^{1/4}$ and $R^{-2}$ 
structural components which dominate the $R <$ 30~kpc and $R >$ 30~kpc regions
of the galaxy, respectively.}.  We propose that this new component is in fact 
the stellar halo of M31 (see also Chapman et~al. 2006).  Taken together, these 
results clarify that the actual 
structural disparities between the Milky Way and M31 may not lie in the 
properties of their halo populations, but rather in the relative sizes of 
the bulges of the two systems.  The discovery of this metal-poor 
stellar halo also provides a powerful confirmation of galaxy formation 
models (see \S\,\ref{m31vsgalaxy}).  

\begin{figure}
\begin{center}
\leavevmode
\includegraphics[width=8cm]{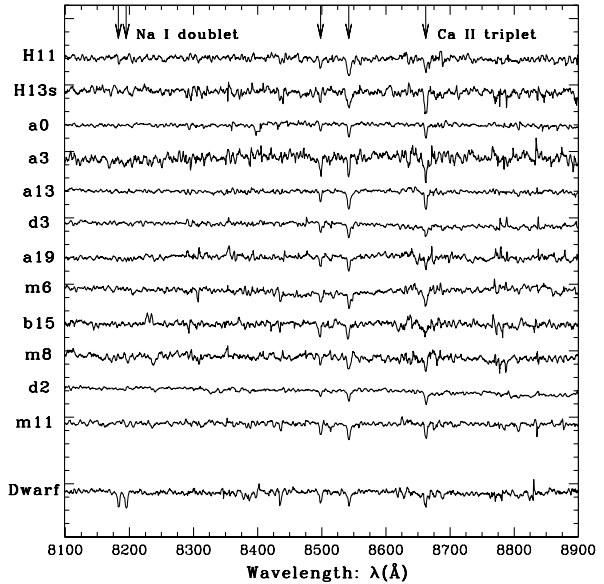}
\end{center}
\caption{Sample M31 RGB spectra, one star from each of our twelve fields, as 
well as the spectrum of a typical foreground Milky Way dwarf star (bottom).
The spectra have been corrected 
to zero velocity, normalized at $\rm\lambda\sim8500\AA$, offset in $y$ 
arbitrarily, and smoothed using a 10~pixel ($\rm3\AA$) boxcar function for 
illustration purposes only.  The locations of the Ca\,{\sc ii} triplet: 
8498, 8542, and $\rm8662\AA$ and pressure sensitive Na\,{\sc i} doublet: 
8183 and $\rm8195\AA$ are indicated with arrows at the top of the figure.  
As expected, the Na\,{\sc i} doublet is measured to be very strong in cool 
foreground Milky Way dwarf stars, but is absent in M31 RGB stars.}
\label{fig:spectra}
\end{figure}

\section{Observations and Data Reduction}\label{data}

The locations of the imaging/spectroscopic fields presented in this study are 
shown in Figure 1.  The wide-field imaging observations were obtained with the 
Kitt Peak National Observatory (KPNO) 4-m telescope and Mosaic camera for most 
of our fields \citep{ostheimer}.  These are shown as gridded squares.  The two 
larger open squares show the positions of very wide-field imaging observations 
obtained with the Canada-France-Hawaii 3.6-meter telescope and MegaCam camera.  
The filled circles indicate the positions of our twelve spectroscopic fields, 
all obtained with the Keck~II telescope and DEIMOS instrument.  The imaging for 
the outer fields was obtained by \cite{ostheimer} 
using the KPNO Mosaic camera, in the Washington System $M$ and $T_2$ bands 
and the intermediate-width $DDO51$ filter \citep{majewski}.  This filter allows 
measurement of the surface-gravity sensitive Mgb and MgH stellar absorption 
features and therefore provides a means to discriminate foreground Milky Way contaminants 
from M31 RGB stars.  The innermost two fields in our study, H11 and H13s, are near 
enough to M31 that there is minimal foreground Milky Way dwarf contamination.  We therefore 
did not observe these fields in the $DDO51$ filter.  Rather, we obtained very wide-field 
Canada-France-Hawaii Telescope (CFHT) photometry in the $g'$ and $i'$ filters using 
the 1~degree$^2$ MegaCam camera.  The $g'$ and $i'$ photometry was converted to 
Johnson-Cousins $V$ and $I$ using standard Landolt field observations.


\begin{table}
\begin{center}
\caption{}
\begin{tabular}{llcccc}
\hline
\hline
\multicolumn{1}{c}{Date} & \multicolumn{1}{c}{Mask} & 
\multicolumn{2}{c}{Pointing center:} & \multicolumn{1}{c}{Field PA} & 
\multicolumn{1}{c}{No. Sci.}  \\ 
& \multicolumn{1}{c}{} &\multicolumn{1}{c}{$\alpha_{\rm J2000}$} &
\multicolumn{1}{c}{$\delta_{\rm J2000}$} &\multicolumn{1}{c}{($^\circ$E of N)} & 
\multicolumn{1}{c}{Targets\tablenotemark{1}} \\ 
& \multicolumn{1}{c}{} & \multicolumn{1}{c}{($\rm^h$:$\rm^m$:$\rm^s$)} &
\multicolumn{1}{c}{($^\circ$:$'$:$''$)} & & \\
\hline
2002 Aug 16  & a3\_1     & 00:48:21.16  & +39:02:39.2 & $+64.2$    & 85   \\
2002 Aug 16  & a0\_1     & 00:51:51.32  & +39:50:21.4 & $-17.9$    & 89   \\
2002 Oct 11  & a3\_2     & 00:47:47.24  & +39:05:56.3 & $+178.2$   & 80   \\
2002 Oct 12  & a0\_2     & 00:51:29.59  & +39:44:00.8 & $+90.0$    & 89   \\
2003 Sep 30  & a13\_1    & 00:42:58.34  & +36:59:19.3 & $+0.0$     & 80   \\
2003 Sep 30  & a13\_2    & 00:41:28.27  & +36:50:19.2 & $+0.0$     & 71   \\
2003 Sep 30  & m11\_1    & 01:29:34.44  & +34:13:45.4 & $+0.0$     & 72   \\ 
2003 Oct 1   & m11\_2    & 01:29:34.35  & +34:27:45.5 & $+0.0$     & 68   \\
2003 Oct 1   & m6\_1     & 01:09:51.75  & +37:46:59.8 & $+0.0$     & 75   \\
2003 Oct 26  & a3\_3     & 00:48:23.17  & +39:12:38.5 & $+270.0$   & 83   \\
2004 June 17 & a0\_3     & 00:51:50.46  & +40:07:00.9 & $+0.0$     & 90   \\
2004 Sep 20  & H11\_1    & 00:46:21.02  & +40:41:31.3 & $+21.0$    & 139  \\
2004 Sep 20  & H11\_2    & 00:46:21.02  & +40:41:31.3 & $-21.0$    & 138  \\ 
2004 Sep 20  & H13s\_1   & 00:44:14.76  & +39:44:18.2 & $+21.0$    & 134  \\
2004 Sep 20  & H13s\_2   & 00:44:14.76  & +39:44:18.2 & $-21.0$    & 138  \\ 
2005 Jun 9   & m6\_2     & 01:08:36.22  &  +37:28:59.6 & $+0.0$    & 72   \\
2005 Jul 7   & m8\_1     & 01:18:11.56  &  +36:16:24.9 & $+0.0$    & 56   \\
2005 Jul 7   & m8\_2     & 01:18:35.87  &  +36:14:30.9 & $+0.0$    & 59   \\
2005 Jul 8   & m11\_3    & 01:30:01.53  &  +34:13:45.4 & $+0.0$    & 80   \\
2005 Jul 8   & m11\_4    & 01:30:37.33  &  +34:13:27.4 & $+0.0$    & 75   \\
2005 Aug 29  & a19\_1    & 00:38:16.05  &  +35:28:07.2 & $-90.0$   & 71   \\
2005 Sep 6   & d2\_1     & 01:17:07.46  &  +33:29:25.1 & $-90.0$   & 139  \\
2005 Sep 6   & d2\_2     & 01:16:43.29  &  +33:34:25.8 & $+0.0$    & 141  \\
2005 Sep 7   & b15\_1    & 00:53:23.63  &  +34:37:16.0 & $-90.0$   & 65   \\
2005 Sep 7   & b15\_3    & 00:53:37.77  &  +34:50:04.1 & $-90.0$   & 74   \\
2005 Sep 8   & d3\_1     & 00:36:03.83  &  +36:27:27.4 & $+90.0$   & 120  \\
2005 Sep 8   & d3\_2     & 00:35:39.61  &  +36:21:41.8 & $+0.0$    & 122  \\

\hline
\end{tabular}
\tablenotetext{1}{Some targets were observed on multiple masks.}
\label{table:masks1}
\end{center}
\end{table}

Details on the imaging observations, slitmask design, spectroscopic observations, 
and data reduction of the fields a0, a3, a13, d3, a19, m6, b15, m8, d2, 
and m11 are given in \S\,2 of \cite{gilbert} and \cite{guh05b}.  Similar information 
for the two innermost fields H11 and H13s can be found in \S\,3 of \cite{kalirai}.  
Keck/DEIMOS spectroscopic observations were obtained in each of the above pointings as 
discussed in \cite{gilbert} and \cite{kalirai}.  We briefly summarize the information 
presented in these papers.  We used the 1200~lines~mm$^{-1}$ grating 
(dispersion = $\rm0.33\AA$~pixel$^{-1}$), providing a spectral resolution 
of $\rm1.3\AA$ (FWHM) for typical 0$\farcs$8 FWHM seeing.  We targeted 
the brightest M31 RGB stars ($20.0<I_0<22.5$ -- the brightest in this window 
may in fact be asymptotic giant branch stars) in the photometry and built 
masks in an iterative process that maximized the total number of slits selected 
for highest priority objects.  
These, in the case of the outer fields, were largely based on the 
position of the star in the CMD and the position in the ($M - DDO51$) versus ($M-T_2$) 
color-color diagram (e.g., Majewski et al. 2000; Palma et al. 2003).  
As mentioned earlier, the latter provides a preselection to ensure high probability 
RGB stars, which is especially important in the outer regions of M31 where true 
RGB stars have sparse density.  For the inner fields, 
a combination of the magnitudes and a stellarity (star-like) limit based on the morphology 
of the sources from SExtractor \citep{bertin} was used to prepare the masks.  In general, 
the high density of objects in the inner fields allowed us to target $\sim$150 targets 
on each mask.  All of the spectra were inspected visually and assigned a quality control 
index based on the quality and number of absorption lines visible.  As discussed in 
\cite{kalirai}, a significant number of the fainter targets yielded spectra that 
did not show any obvious features due to low signal-to-noise (S/N).  We determine 
the velocities for all objects that contain at least two 
spectral features (at least one definite and one other marginal line) by 
cross-correlating the observed spectra with a large database of template stellar and 
emission- and absorption-line galaxy spectra.  The mean S/N of these 
{\it good\/} spectra is $\sim$10 per pixel and the velocity uncertainty from the 
cross-correlation is empirically estimated to be $\sim$15 km~s$^{-1}$.  The mean magnitude 
for these stars (with S/N $\sim$ 10 per pixel) is $I_0 \sim 21.2$.

After removing galaxies, the twelve fields (27 masks) in this study contain 
1070 stars for which we obtained a reliable velocity.  Figure \ref{fig:spectra} 
shows a single representative spectrum of an M31 RGB star from each of these fields.  
Table 1 presents a summary of the observations.  

\section{A Clean Sample of M31 RGB Stars}\label{cleansample}

\begin{figure*}[ht]
\begin{center}
\leavevmode
\includegraphics[width=11cm]{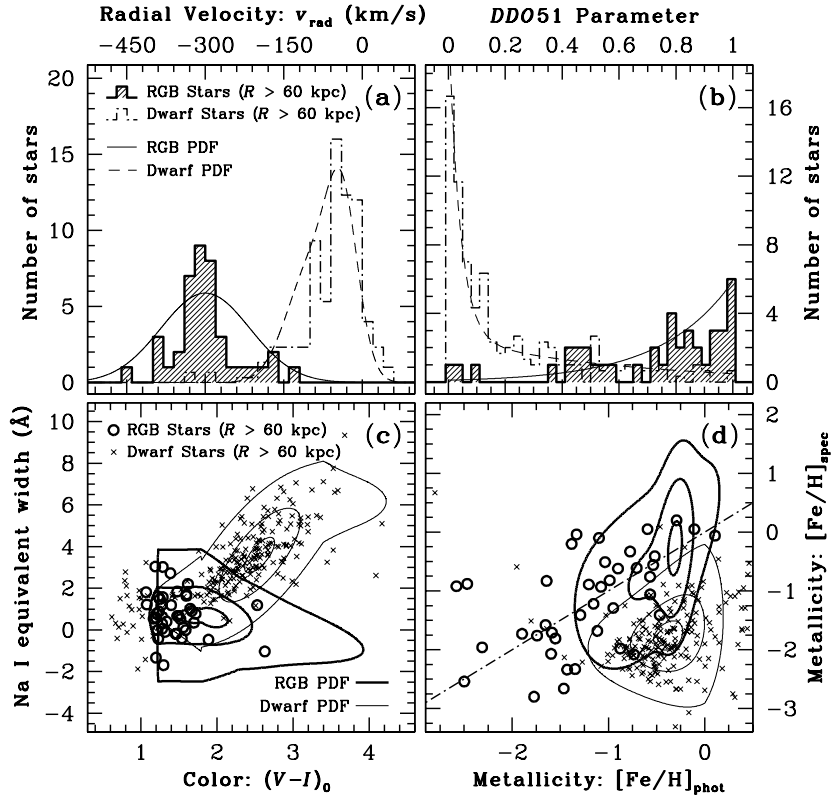}
\end{center}
\caption{Properties of confirmed M31 RGB stars in six of our outer spectroscopic 
fields (a13, a19, m6, b15, m8, and m11 - all located at $R >$ 60 kpc).  Four of the five 
diagnostics used to distinguish RGB stars from foreground Galactic dwarf star contaminants 
are illustrated. (a) Radial velocity distribution of M31 RGB stars and dwarf stars 
[shaded-bold and dot-dashed histograms, respectively], compared
to empirical probability distribution functions (PDFs) derived from training sets
consisting of definite RGB and dwarf stars in all of our fields (solid and dashed 
curves, respectively).  (b) Same as (a) for the $DDO51$ parameter, which is based on the
star's location in ($M-DDO51$) vs. ($M-T_2$) color-color space.  For both (a) and (b), 
we have scaled back the dwarf star histograms by a factor of three for clarity and 
arbitrarily scaled the training set PDFs. (c) Equivalent width of the Na\,{\sc i} $\rm8190\AA$ 
absorption band versus de-reddened ($V-I$)${_0}$ color for the RGB and dwarf 
stars [bold circles and crosses, respectively]. Bold and thin contours show 
10\%, 50\%, and 90\% enclosed fractions for RGB and dwarf star PDFs, respectively. 
(d) Same as (c) for the photometric metallicity estimate (CMD based, see \S\,\ref{photmet}) 
versus spectroscopic metallicity estimate (based on the strength of the $\rm8500\AA$ Ca\,{\sc ii}
triplet, see \S\,\ref{specmet}). The dot-dashed diagonal line shows the one-to-one relation, 
which is followed nicely by confirmed RGB stars. It is reassuring that the M31 RGB 
distribution generally follows the RGB PDF and not the dwarf PDF.}
\label{fig:pdfs}
\end{figure*}

A fraction of the 1070 stars in our sample are in fact foreground Milky Way 
dwarfs.  In order to measure the MDF in M31's bulge and halo, we first need to 
isolate M31 RGB stars from this 
contamination.  We have developed a sensitive technique that uses probability 
distribution functions (PDFs) calculated from a training set of known RGB and dwarf 
stars to provide this discriminant \citep{gilbert}.  Using five criteria --- 
(1)~radial velocity, (2)~intermediate-width $DDO51$ photometry, (3)~Na\,{\sc i} 
equivalent width (surface gravity sensitive), (4)~position in the CMD, and 
(5)~comparison between photometric and spectroscopic [Fe/H] estimates --- 
each star in our sample is assigned five likelihood values of being a 
giant or a dwarf based on its location within each diagnostic plot.  The individual 
probabilities are then combined to yield the final discriminant of whether the 
star is truly an M31 RGB star.  In Figure \ref{fig:pdfs} we show the properties 
of confirmed M31 RGB stars in six of our outer spectroscopic fields, a13, a19, 
m6, b15, m8, and m11.  We have plotted four of the five diagnostics used to distinguish 
RGB stars from dwarfs, and have also overlayed empirical PDFs derived from 
the training set of definite RGB (solid darker curve) and dwarf (dashed/thin curve) 
stars.  It is clear that the outer RGB stars predominantly agree with the 
RGB training set in {\it all\/} of their properties.

\begin{figure}
\begin{center}
\leavevmode
\includegraphics[width=8cm]{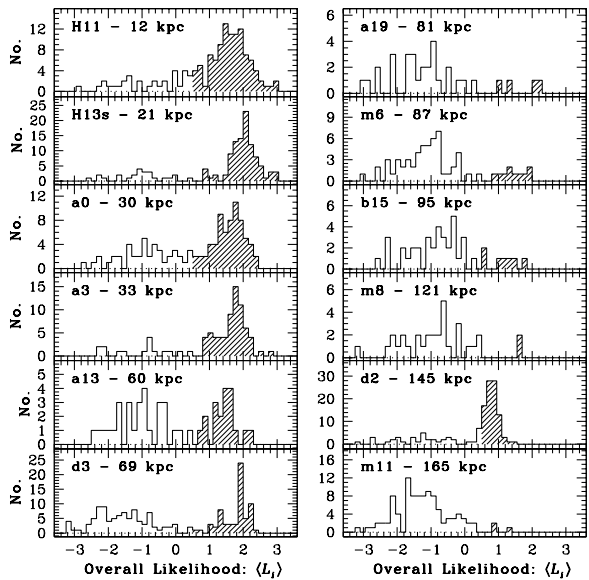}
\end{center}
\caption{The weighted average likelihood distinguishing M31 RGB stars ($\langle$$L_i$$\rangle$ 
$>$ 0.5 -- shaded) from Milky Way dwarfs for each of our twelve fields.  The fields 
are arranged in order of increasing projected distance from M31.}  
\label{fig:lhist}
\end{figure}

A detailed discussion of the procedure used to average the individual PDFs into a likelihood 
distribution is given in \cite{gilbert}.  In Figure~\ref{fig:lhist} we present the 
final distribution of weighted average likelihoods, $\langle$$L_i$$\rangle$.  Although 
any star with $\langle$$L_i$$\rangle$ $>$ 0 is a preferred M31 RGB star (whereas stars with 
$\langle$$L_i$$\rangle$ $<$ 0 are likely to be Milky Way dwarfs), we only select stars 
with $\langle$$L_i$$\rangle$ $>$ 0.5 for our clean M31 RGB sample.  This ensures that the 
star is three times more likely to be an M31 RGB member than a Milky Way dwarf 
(see Gilbert et al. 2006).  We also impose a strict cut to eliminate a few very 
blue stars that are inconsistent with the color of M31's RGB (see Gilbert et 
al. 2006).  The remaining sample (530 stars) are shown as shaded histograms in 
Figure~\ref{fig:lhist}.  We stress that although it may not be clear from any 
individual diagnostic whether or not a star is a dwarf or a giant, the combination 
of the ten PDFs presents an unambiguous answer.  


\begin{table*}
\begin{center}
\caption{}
\begin{tabular}{lcccccl}
\hline
\hline
\multicolumn{1}{c}{Mask} & \multicolumn{1}{c}{$R$} & \multicolumn{1}{c}{No. Sci.}  & \multicolumn{1}{c}{No. M31 RGB} & 
\multicolumn{1}{c}{No. Milky Way} & \multicolumn{1}{c}{No. M31 RGB} & \multicolumn{1}{c}{Comments} \\ 
& \multicolumn{1}{c}{(kpc)} & \multicolumn{1}{c}{Targets\tablenotemark{1}} & 
\multicolumn{1}{c}{Stars\tablenotemark{2}} & \multicolumn{1}{c}{Dwarfs\tablenotemark{2}} & \multicolumn{1}{c}{Spheroid Stars}\\
\hline
H11    & 12  & 277 & 106 & 18 & 106 &\\
H13s   & 21  & 272 & 104 & 20 & 21 & giant southern stream field \\
a0     & 30  & 268 & 67  & 30 & 67 & \\
a3     & 33  & 248 & 68  & 12 & 20 & giant southern stream field \\
a13    & 60  & 151 & 18  & 23 & 18 & \\
d3     & 69  & 242 & 60  & 89 & 6  & And III dwarf spheroidal field \\
a19    & 81  & 71  & 4   & 23 & 4  & \\
m6     & 87  & 147 & 9   & 49 & 9 & \\
b15    & 95  & 139 & 6   & 25 & 6  & \\
m8     & 121 & 115 & 1   & 24 & 1  & \\
d2     & 145 & 280 & 84  & 38 & 0  & And II dwarf spheroidal field \\
m11    & 165 & 295 & 3   & 82 & 3  & \\

\hline
\end{tabular}
\tablenotetext{1}{Some targets were observed on multiple masks.} 
\tablenotetext{2}{Numbers given indicate {\it secure\/} M31 RGB stars 
($\langle$$L_i$$\rangle$ $>$ 0.5) and Milky Way dwarfs 
($\langle$$L_i$$\rangle$ $<$ $\rm-$0.5) - see \S\,\ref{cleansample} and 
Gilbert et al. (2006).} 
\label{table:masks2}
\end{center}
\end{table*}


\begin{figure}
\begin{center}
\leavevmode
\includegraphics[width=8cm]{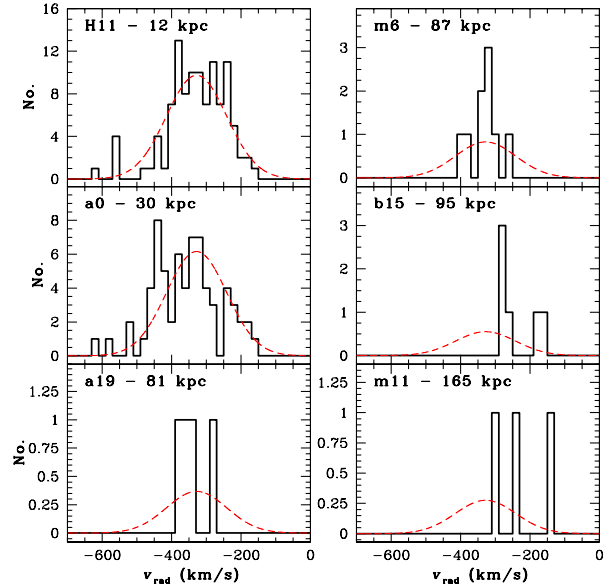}
\end{center}
\caption{Radial velocity histograms for confirmed M31 RGB stars in six representative 
bulge and halo fields, arranged in order of increasing distance from M31.  A scaled 
Gaussian with the characteristic velocity and dispersion of our most populous field, H11, 
has been overlain on the data to show that the distribution of stars in each of these 
fields is {\it roughly\/} consistent with a broad Gaussian centered near M31's 
systemic velocity.}
\label{fig:kinematics}
\end{figure}

As a point of interest, we note that some other studies of M31 have isolated 
Milky Way dwarfs from M31 RGB stars based solely on a velocity cut.  Using our PDFs 
for the two minor axis fields H11 and a0, we calculate that such a cut will 
introduce a significant amount of Milky Way dwarf star contamination into the 
putative RGB sample.  For example, invoking only a velocity cut of $v_{\rm rad} <$ 
$\rm-$100~km~s$^{-1}$ to isolate M31 RGB stars from Milky Way dwarfs will lead 
to a contamination fraction of $\sim$15\% in our data.  Similarly, for a 
velocity cut of $v_{\rm rad} <$ $\rm-$150~km~s$^{-1}$ cut, over 10\% of all 
classified M31 RGB stars would actually be Milky Way dwarfs.  For fields that 
sample the extended disk of M31 in the N-E or N-W quadrants, the contamination 
rates could be higher due to the rotation of M31's disk which would cause the 
velocity histogram to overlap with the Milky Way dwarf distribution.  This would 
of course be offset to some degree (perhaps completely) depending on the 
distance of these fields from M31's center (i.e., the density of M31 RGB stars 
is much higher along the major axis).

In Table 2, we present the numbers of stars in each field that were determined to be 
{\it secure\/} M31 RGB stars and Milky Way dwarfs (measured using a 
$\langle$$L_i$$\rangle$ $>$ 0.5 cut) from the procedure outlined above.  Further 
analysis from this point will involve only the cleaned M31 RGB sample.


\section{Kinematics of Sample} \label{kin:section}

In this section we briefly present kinematics of M31 RGB stars in some of our fields and 
demonstrate that the data are consistent with a broad Gaussian centered at M31's 
systemic velocity.  As can be seen from Figure~\ref{fig:fields}, five of our pointings 
are located on the minor axis (H11, a0, m6, m8, and m11), two of the pointings are 
located on the giant southern stream (H13s and a3), and five pointings are removed from 
the minor axis (a13, d3, a19, b15, and d2).  The data for the minor axis fields and the 
off axis pointings are likely fair representations of the bulge and halo of M31 
(see \S\,\ref{definebulgehalo}).  However, the targeted stars in the giant southern stream (H13s and 
a3) and dwarf satellites (d2--And II and d3--And III) have very different kinematics 
than hot bulge or halo stars.  As demonstrated in \cite{ibata04}, \cite{guh05b}, and 
\cite{kalirai}, the stream consists of a kinematically cold population of 
stars blueshifted relative to M31 (in H13s stream stars have $v_{\rm rad} <$ $\rm-$380~km~s$^{-1}$, 
in a3 stream stars have $v_{\rm rad} <$ $\rm-$400~km~s$^{-1}$).  The stream is well removed, 
both azimuthally and radially, from all of our other pointings and does not contaminate 
these fields.  Similarly, d2 and d3 are dominated by stars belonging to the two dwarf 
satellites And II ($\rm-$230~km~s$^{-1}$ $< v_{\rm rad} <$ $\rm-$150~km~s$^{-1}$) and 
And III ($\rm-$390~km~s$^{-1}$ $< v_{\rm rad} <$ $\rm-$325~km~s$^{-1}$).  For each 
of these four fields we easily remove the contribution of the stream and satellite 
galaxies by making these velocity cuts, leaving a pure M31 RGB bulge and halo sample.

Figure~\ref{fig:kinematics} presents the velocity histograms for six fields in M31 that 
are not contaminated by the giant southern stream or one of the dwarf satellites.  
Velocity histograms for the giant southern stream can be found in \cite{kalirai} (H13s) 
and \cite{guh05b} (a3).  The fields shown here 
have been randomly chosen and sample the entire distance range over which we measure 
M31 RGB stars.  The two minor axis fields with a large sample of stars, H11 and a0, 
clearly demonstrate that the majority of these stars form a broad, hot component 
centered near M31's systemic velocity. For our most populated field (H11), we find 
$v_{\rm rad} \sim$ $\rm-$330~km~s$^{-1}$ with a dispersion of $\sigma_v \sim 90$~km~s$^{-1}$ 
(skew = 1.3).  The velocity histograms for our outermost fields are affected by small number 
statistics.  However, almost all confirmed M31 RGB stars in these fields (e.g., a19, m6, b15, 
and m11) have kinematics consistent with M31's systemic velocity, $v_{\rm rad} \sim$ 
$\rm-$300~km~s$^{-1}$.  To demonstrate this, we have overlaid a scaled Gaussian with the 
mean and dispersion of the H11 field in each of these panels (dashed curve).

The combination of Figures~\ref{fig:lhist} and \ref{fig:kinematics} shows that 
we have detected a population of RGB stars belonging to M31's bulge and halo in eleven of 
our twelve fields (the exception being field d2, which is dominated by stars belonging to M31's 
dwarf satellite, And II).  The final starcounts for these populations, in each field, are given 
in column 6 of Table 2 (this is the number after eliminating foreground dwarfs, M31 stream, 
and M31 satellite galaxy stars). Adding up all of the data, our sample consists of 
261 bona fide M31 RGB bulge and halo stars, 47 of which have $R >$ 60~kpc.  We stress that 
prior to this work, very few (if any) M31 halo RGB stars have been spectroscopically 
confirmed at these large radii.


\section{Metallicity Measurements}

We determine the metallicities of stars in our sample using two independent 
techniques, photometrically from the ($I$, $V{\rm-}I$) CMD and spectroscopically 
from the Ca\,{\sc ii} triplet ($\rm\lambda\sim8500\AA$).

\begin{figure}
\begin{center}
\leavevmode
\includegraphics[width=8cm]{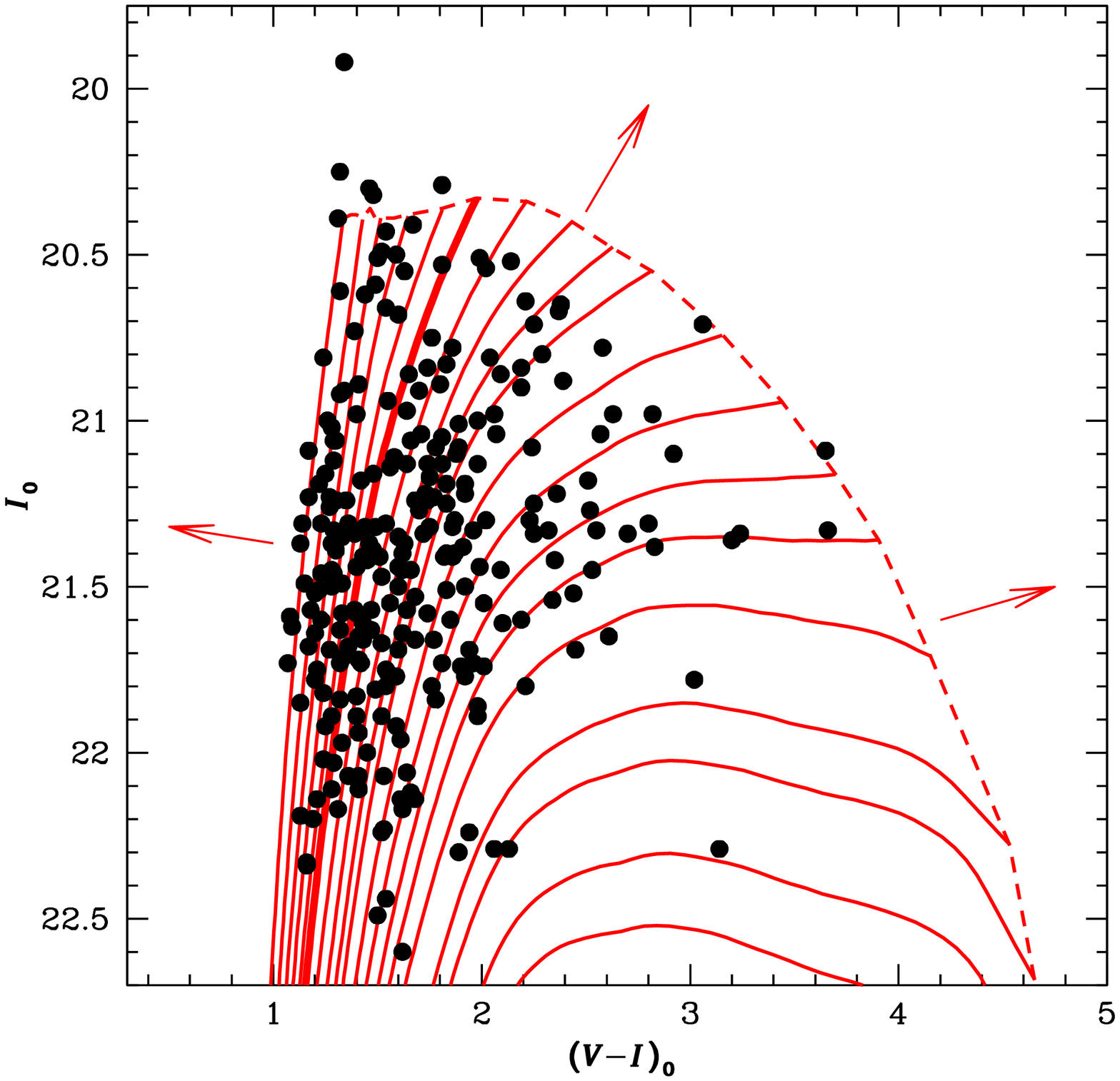}
\end{center}
\caption{CMD of all M31 RGB bulge and halo stars.  Most stars are confined within a grid 
of theoretical isochrones \citep{vandenberg} of age 12~Gyr spanning the metallicity range, 
[Fe/H] = $\rm-$2.31 to $+$0.49 (bold isochrone: [Fe/H] = $\rm-$1.0).  A few outliers 
are found above the tip of the RGB (dashed line).  The arrows indicate the direction in 
which we extrapolated metallicity measurements for a few stars.  Section \ref{photmet} discusses 
the determination of these metallicities for this sample.}
\label{fig:CMD}
\end{figure}


\subsection{Photometric Metallicity Determination}\label{photmet}

In Figure~\ref{fig:CMD} we present the CMD for our entire M31 RGB bulge and halo 
sample.  For this purpose, we first converted the Washington System ($M$, $T_2$) 
photometry into Johnson-Cousins ($I$, $V{\rm-}I$) magnitudes using the relations in 
\cite{majewski}.  The systematic error in the slope of the color conversion from 
these relations is less than 3.6\%, as measured by \cite{majewski}.  Also shown are 
several theoretical isochrones ranging in metallicity from [Fe/H] = $\rm-$2.31 -- $+$0.49, 
all with an age of 12 Gyr and [$\alpha$/Fe] = 0.0 \citep{vandenberg}.  These have been 
adjusted to a distance of 783~kpc ($(m-M)_0=24.47$).  Most of our sample is indeed confined 
within the bounds of the most metal-rich and metal-poor isochrones on the CMD.  For reference, 
the bolder isochrone has a metallicity of [Fe/H] = $\rm-$1.0.  The dashed line at the top 
of the isochrones indicates the tip of the RGB.  

We compute photometric metallicities ([Fe/H]$_{\rm phot}$) for these 
stars by first measuring the length in the CMD of a segment that extends from the most 
metal-poor isochrone to the most metal-rich.  This is done at 35 points along the 
set of isochrones, extending from the base of our RGB sample in Figure~\ref{fig:CMD} to 
well above the tip of the RGB (by linearly extrapolating the isochrones in brightness and 
color).  The dashed curve on Figure~\ref{fig:CMD} represents one of these length 
segments.  We then normalize these 35 segments by their total length, hence computing 
an index, $X$, equal to the fractional distance that a point is away from the most metal-poor 
isochrone.  This $X$ parameter is now a smooth function of metallicity, for a given $Y$ range, 
the distance above the base of the RGB in our sample (extending from 0 at the base to 1 at 
the tip of the RBG).  Photometric metallicities are derived by measuring each star's $X$ 
position and interpolating that within the relation that is appropriate given its $Y$ value.  
Although we do this for many $Y$ ranges (bins of size 0.1), we find that the $X$ versus [Fe/H] 
relations are not very sensitive to this parameter and vary smoothly and slowly over the 
entire magnitude range of RGB stars in our sample.  Metallicities for confirmed RGB 
stars that lie outside the range of the isochrones are derived by extrapolating the $X$ 
versus [Fe/H] relations.  As the CMD in Figure~\ref{fig:CMD} shows, this mild extrapolation 
was required for only a few stars.

We also tested our [Fe/H]$_{\rm phot}$ measurements derived above using two 
independent sets of isochrones.  Interpolating metallicities within a grid of the Padova 
models \citep{girardi} and the Yale-Yonsei models (Y$^2$ - Demarque et al. 2004) give consistent 
values to those derived using the \cite{vandenberg} models.  For a typical star, the mean [Fe/H] 
varies by less than 0.15 dex depending on which isochrone set is adopted.  We find that the 
Padova models systematically yield slightly more metal-rich values than both the 
\cite{vandenberg} and \cite{demarque} models.  We choose to adopt the \cite{vandenberg} models 
as they also provide a grid of $\alpha$-enhanced isochrones over a broad metallicity range (see 
\S\,\ref{mdftrends}).

\begin{figure}
\begin{center}
\leavevmode
\includegraphics[width=8cm]{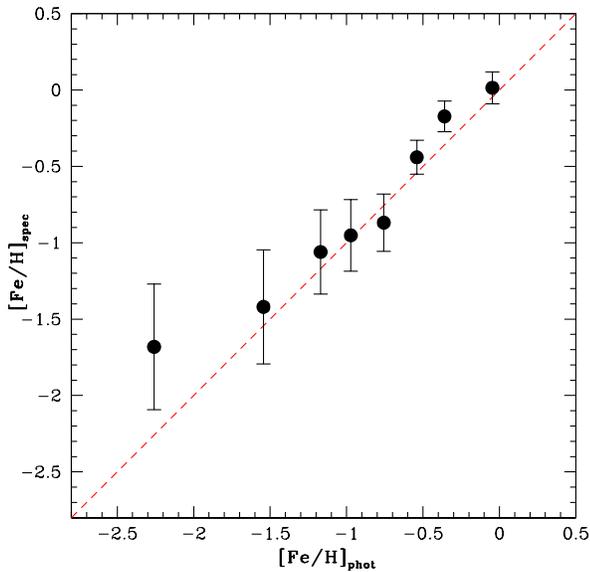}
\end{center}
\caption{Comparison of spectroscopic ([Fe/H]$_{\rm spec}$) versus photometric
([Fe/H]$_{\rm phot}$) 
metallicities of all M31 RGB bulge and halo stars.  We have used a minimum bin 
size of 0.2 dex in [Fe/H]$_{\rm phot}$ to bin the sample, while 
ensuring a minimum of 20 stars in each bin. The data span a wide metallicity range and 
the two independently measured quantities are found to be in good agreement 
with one another over most of this range.  For reference, a 1:1 relation is shown 
as a dashed line.}
\label{fig:metcomp}
\end{figure}

\subsection{Spectroscopic Metallicity Determination}\label{specmet}

Independent of the method discussed above, we also determine metallicities for all of our 
M31 RGB bulge and halo stars using their spectra.  This procedure relies on measuring the 
equivalent widths of the three Ca\,{\sc ii} absorption lines 
(see Figure~\ref{fig:spectra}).  The strengths of each of these lines are combined 
to yield a reduced equivalent width according to the prescription described in 
\cite{rutledge97a}.  This reduced equivalent width is then calibrated 
empirically based on Galactic globular cluster RGB stars to yield [Fe/H]$_{\rm spec}$ 
\citep{rutledge97b}.  Further details on this procedure are given in \cite{gilbert} 
and \cite{guh05b}.

In Figure~\ref{fig:metcomp}, we present a comparison of the two independently determined 
metallicity measurements for our sample.  Given the larger scatter in [Fe/H]$_{\rm spec}$, we 
have binned our sample by using a minimum bin size of 0.2 dex in [Fe/H]$_{\rm phot}$ while 
ensuring $>$20 stars in each bin.  The individual [Fe/H]$_{\rm phot}$ and binned [Fe/H]$_{\rm spec}$ 
measurements are found to be in good agreement with one another over most of the metallicity 
range (1:1 line is shown), indicating that there are no systematic variations 
in our metallicity scales.  This agrees with the results of \cite{reit02} who 
also find a nice agreement between photometric and spectroscopic [Fe/H] measurements for 
RGB stars.  

\begin{figure}
\begin{center}
\leavevmode
\includegraphics[width=8cm]{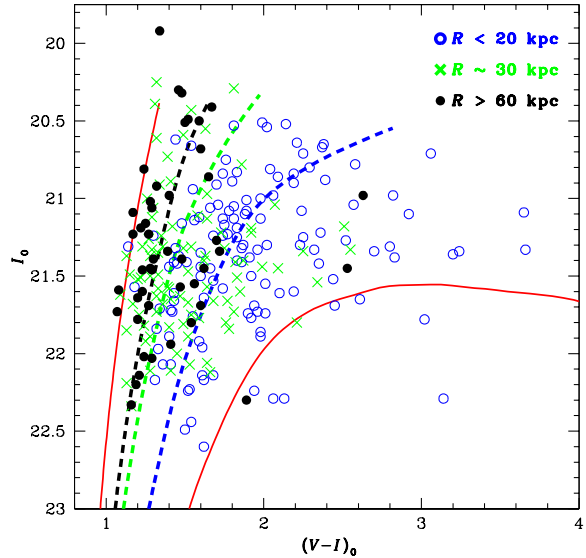}
\end{center}
\caption{The distribution of M31 RGB stars in the bulge ($R <$ 20~kpc, open circles), crossover region 
($R \sim$ 30~kpc, crosses), and halo ($R >$ 60~kpc, filled circles) on the CMD show differences.  In 
\S\,\ref{mdftrends}, we provide evidence that these differences are related to the MDF 
of stars in M31's different structural components.  Solid curves represent theoretical 
isochrones \citep{vandenberg} with [Fe/H] = $\rm-$2.31 (left) and 0.0 (right).  The dashed curves 
are discussed in \S\,\ref{mdftrends}.}
\label{fig:3cmds}
\end{figure}

\bigskip

We also note here that most classes of systematic biases/errors in our measurement 
of [Fe/H]$_{\rm phot}$ and [Fe/H]$_{\rm spec}$ (e.g., age errors, systematic errors in 
the models, etc.) will tend to affect different fields similarly.  Therefore, although 
the absolute [Fe/H]$_{\rm phot}$ for any given star may have a large error 
(e.g., due to photometric calibration or incorrect distance modulus), the relative 
comparison of [Fe/H]$_{\rm phot}$ between different fields within our sample is 
more accurate.  In \S\,\ref{biases} we discuss several possible biases. 

\begin{figure*}[ht]
\begin{center}
\leavevmode
\includegraphics[width=11cm, angle=270]{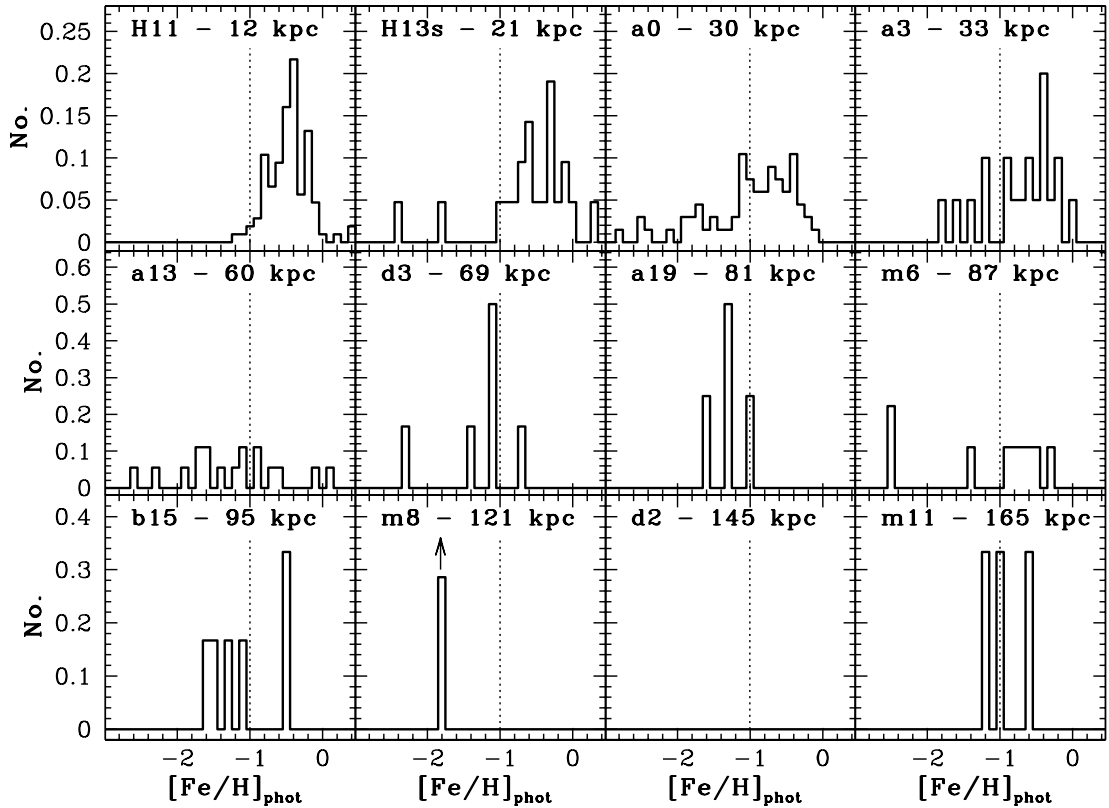}
\end{center}
\caption{Normalized MDFs of confirmed M31 RGB bulge and halo stars in each of our 
twelve fields.  A general trend indicating the presence of metal-poor stars in the halo of M31 
is seen.  As a guide, the dotted line is fixed at [Fe/H] = $\rm-$1.0 in all panels.  
An arrow is indicated in the panel for m8 as we have re-normalized the MDF for 
this field.  The cumulative distributions corresponding to these twelve fields 
are shown in Figure~\ref{fig:cummdfphot}.  In Figure~\ref{fig:3bins} we 
bin all of the halo data ($R >$ 60~kpc) together to produce a more statistically 
significant sample from which we measure the mean metallicity and dispersion of 
M31's halo.  Note: field d2 does not contain any M31 RGB stars that are not 
members of And II.}
\label{fig:mdfphot}
\end{figure*}

\section{Analysis}

\subsection{The Bulge and Halo Samples} \label{definebulgehalo}

As Table 2 shows, our M31 coverage ranges from a field located at $R$ = 12~kpc (H11) 
to a field located at $R$ = 165~kpc (m11).  Given this very large range in radius from 
M31's center, and our small number statistics in the outermost fields, we separate our 
fields into three broad regions.  We define the bulge sample to be that represented by 
the H11 and H13s fields, the crossover region by the a0 and a3 fields, and the halo 
by the a13, d3, a19, m6, b15, m8, d2, and m11 fields (d2 containing no halo members).  
This separation is justified 
naturally for several reasons.  First, as discussed in \S\,\ref{intro}, several authors 
have shown that there is no observed metallicity gradient in M31 out to $R \sim$ 
30~kpc \citep*[e.g.,][]{durrell04}.  One study, Irwin et al. (2006), suggests that there is 
no metallicity gradient out to $\sim$45~kpc.  Second, \cite{guh06a} and \cite{irwin} have 
recently shown that a power-law $R^{-2}$ profile dominates beyond this radius whereas 
a de Vaucouleurs $R^{1/4}$ surface brightness describes regions interior to this 
radius.  \cite{guh06a} further demonstrate that the crossover of M31's bulge and halo 
occurs near this radius.  Therefore, $R \sim$ 30~kpc is a natural choice for separating 
the inner and outer samples.  In addition to the two fields located near this intermediate 
radius (a0 and a3), two of our best sampled fields are located interior to this radius 
(H11 and H13s) while the remaining fields are located beyond this radius.

\subsubsection{Color-Magnitude Diagrams}

Figure~\ref{fig:3cmds} illustrates the CMD for M31 RGB stars in each of the three 
broad radial bins defined above.  Also shown are two theoretical isochrones (solid 
curves - Vandenberg, Bergbusch, \& Dowler 2005) with [Fe/H] = $\rm-$2.31 (left) and 
[Fe/H] = 0.0 (right).  As we showed in Figure~\ref{fig:CMD}, the distribution of 
M31 RGB stars nicely follows the shapes of the isochrones.  However, we can now see that 
some clear differences exist between stars belonging to the inner ($R <$ 20~kpc, 
open circles), crossover ($R \sim$ 30~kpc, crosses), and outer ($R >$ 60~kpc, filled 
circles) regions of M31.  We show quantitatively below that the M31 RGBs in each 
of these regions {\it are\/} significantly different and that these differences 
relate to the MDF of each population.  The dashed curves will be discussed later.


\begin{figure*}
\begin{center}
\leavevmode
\includegraphics[width=11cm, angle=270]{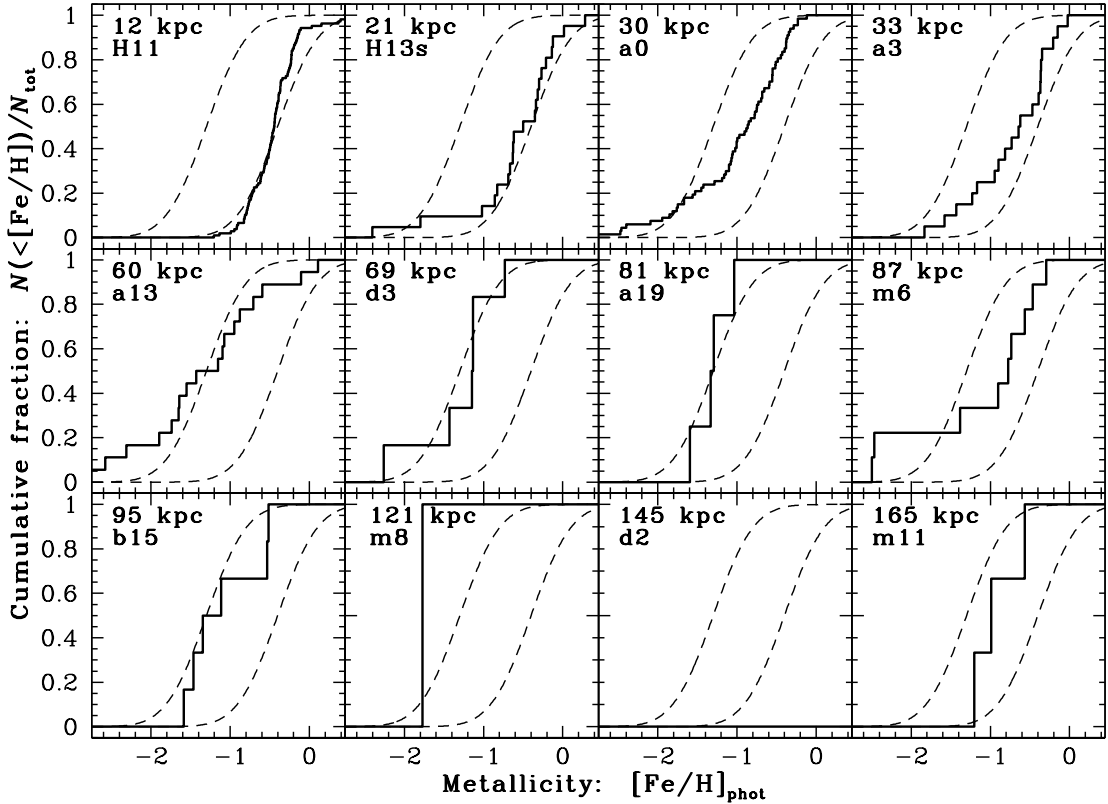}
\end{center}
\caption{Cumulative distributions of [Fe/H]$_{\rm phot}$ for M31 RGB bulge and halo 
stars in each of our twelve fields (solid curves).  The dashed curves represent fixed 
Gaussians with [Fe/H]$_{\rm phot}$ = $\rm-$1.3 and $\rm-$0.4 to be used as guides.  The same 
trend seen in the discrete distributions in Figure~\ref{fig:mdfphot} is confirmed here.
The outer fields are dominated by more metal-poor stars relative to the inner fields.  
Note: field d2 does not contain any M31 RGB stars that are not members of And II.}
\label{fig:cummdfphot}
\end{figure*}

\vskip 1.0cm

\subsection{Metallicity Distribution Function and Radial Trends} \label{mdftrends}

Before presenting our MDFs for confirmed M31 RGB stars in each of the inner, 
crossover, and outer regions defined above, we construct MDFs for each of our 
twelve fields independently.  These are shown in Figure~\ref{fig:mdfphot} along 
with a fixed guide at [Fe/H] = $\rm-$1.0 (dotted line).  The innermost 
fields, H11 and H13s, contain mostly stars with intermediate/metal-rich compositions.  
The a0 and a3 fields at $R \sim$ 30~kpc are dominated by a much broader MDF that 
extends to more metal-poor stars while still containing a metal-rich component.  
This MDF nicely supports the finding in \cite{guh06a} that a 
roughly equal mix of bulge (metal rich) and halo (metal poor) M31 
stars reside at $R \sim$ 30~kpc (see \S\,\ref{bulgehaloratio}).  Although individually 
affected by small number 
statistics, the outermost fields (second and third rows) in our study are dominated 
by a large number of metal-poor stars ([Fe/H]$_{\rm phot}$ $< \rm-$1).  Directly 
comparing these fields to the inner fields, H11 and H13s shows a very obvious 
metallicity gradient in M31.  Less than $\sim$4\% of all stars in the two inner fields 
($R <$ 21~kpc) combined have ([Fe/H]$_{\rm phot}$ $\leq \rm-$1) whereas almost 65\% 
of stars in the outer fields ($R >$ 60~kpc) are this metal poor. Ostheimer (2003) found 
evidence for similar shifts in the proportions of metal rich and metal poor
stars as a function of radius from a purely photometric analysis of all giant star 
candidates in these same fields.

In Figure~\ref{fig:cummdfphot} we present cumulative [Fe/H]$_{\rm phot}$ distributions 
for each of the twelve fields.  The distributions confirm the conclusions drawn above.  
The outermost halo fields contain metal-deficient stars relative to the inner fields.  
As a guide, we have also shown two Gaussian distributions with means at 
[Fe/H]$_{\rm phot}$ = $\rm-$1.3 and $\rm-$0.4 and $\sigma$ = 0.4 (dashed curves).

\begin{figure*}
\begin{center}
\leavevmode
\includegraphics[width=11cm]{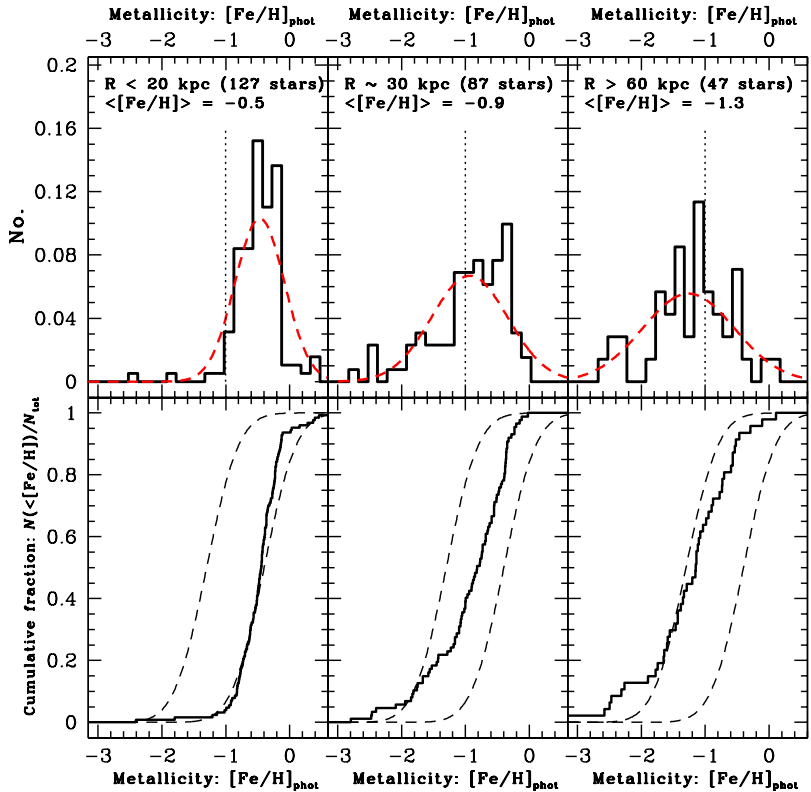}
\end{center}
\caption{({\it Top\/})---Normalized MDFs for each of the bulge (H11 and H13s), crossover 
region (a0 and a3), and halo (a13, d3, a19, m6, b15, m8, d2, and m11) regions. 
The dashed curves represent Gaussians with parameters fixed to the measured mean and 
dispersion of each MDF.  The vertical dotted line is held fixed at [Fe/H] = $\rm-$1.0 
to guide the eye.~~~ ({\it Bottom\/})---Cumulative distributions, corresponding to the 
above panels.  The dashed curves represent fixed Gaussians with mean [Fe/H] = $\rm-$1.3 
and $\rm-$0.4 ($\sigma$ = 0.4).  The data unequivocally show a metallicity gradient 
in M31.}
\label{fig:3bins}
\end{figure*}

In Figure~\ref{fig:3bins} we present both the MDFs (top) and the cumulative 
Gaussians (bottom) for the bulge (H11 and H13s), crossover region 
(a0 and a3), and the halo (a13, d3, a19, m6, b15, m8, d2, and m11) of 
M31.  All histograms have been normalized to unity for clarity.  These data clearly 
show that we have indeed detected a metallicity gradient in M31. For $R <$ 20~kpc, 
we find $\rm\langle[Fe/H]_{\rm phot}\rangle=-0.47\pm0.03$ ($\sigma$ = 0.39), for $R \sim$ 30~kpc 
we find $\rm\langle[Fe/H]_{\rm phot}\rangle=-0.94\pm0.06$ ($\sigma$ = 0.60), and for $R >$ 60~kpc, 
we find $\rm\langle[Fe/H]_{\rm phot}\rangle=-1.26\pm0.10$ ($\sigma$ = 0.72).  Gaussian models 
with the mean and dispersion of the stars comprising the MDF in each of these groups are 
overlaid on the differential histograms as dashed curves.  For completeness, we note that our 
spectroscopic metallicity for the outer halo ($\rm\langle[Fe/H]_{\rm spec}\rangle=-1.24\pm0.12$, 
$\sigma$ = 0.85) is in excellent agreement with the measured photometric metallicity.  

The CMDs for stars belonging to each of these three regions were presented in 
Figure~\ref{fig:3cmds}.  To illustrate the differences in the three CMDs, we have 
overlain isochrones in each panel of Figure~\ref{fig:3cmds} with the approximate mean 
metallicity of the respective population as determined above (dashed curves).

The photometric metallicities given above have been derived assuming isochrones with 
[$\alpha$/Fe] = 0.0.  If we assume that M31's halo is $\alpha$-enhanced, with 
[$\alpha$/Fe] = $+$0.3, the mean metallicity of the outer halo is found to be 
$\rm\langle[Fe/H]_{\rm phot}\rangle=-1.48\pm0.11$ ($\sigma$ = 0.73).


\subsection{Kolmogorov-Smirnov Tests} \label{kstest}

To test whether or not the MDFs of the inner, crossover, and outer regions of M31 
(as defined above) truly differ significantly from one another, we have applied the 
two-sided Kolmogorov-Smirnov (K-S) test \citep{press}.  The K-S test 
makes no assumption about the distribution of the data and is therefore insensitive to  
possible spurious biases from arbitrary binning of data.  The test looks at the cumulative 
fraction of each histogram and returns a statistic, $D$, which measures the maximum vertical 
deviation between the two curves.  This statistic yields a significance level probability, $P$, 
that the null hypothesis (i.e., that the two data sets are drawn from the same distribution) 
is true.  Therefore, small values of $P$ show that the two MDFs are significantly different 
whereas larger values (e.g., $\gtrsim$0.1) indicate that the two MDFs may have been 
drawn from the same distribution.

We find that the $R <$ 20~kpc inner M31 MDF is significantly different from the $R \sim$ 
30~kpc MDF, yielding a tiny probability $P$ = 9.9$\times$10$^{-8}$ of the two being drawn 
from the same parent distribution.  We also find that the $R \sim$ 30~kpc MDF is significantly 
different from the halo MDF, $R >$ 60~kpc.  For this, the two-sided K-S test returns a 
probability of $P$ = 7.5$\times$10$^{-3}$.  Additional tests confirm that the a0 and a3 MDFs 
that have been combined to form the $R \sim$ 30~kpc population are likely drawn from the same 
distribution, $P$ = 0.23 (at least it can not be shown that they are drawn from different 
distributions).  Directly comparing the inner fields, H11 and H13s, with any of the outer 
fields a13, d3, a19, m6, b15, m8, or m11 shows them to be significantly 
different.  Therefore, the K-S tests confirm our earlier conclusion that we have detected 
a metallicity gradient in M31.

\begin{figure}
\begin{center}
\leavevmode
\includegraphics[width=8cm]{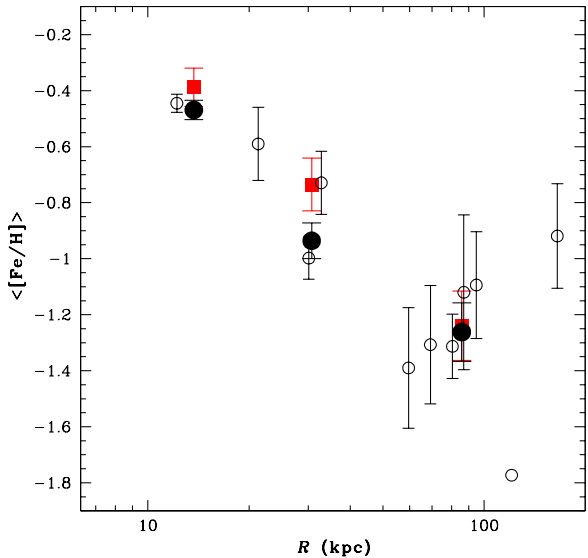}
\end{center}
\caption{The mean metallicity of M31 RGB stars in each of our fields as a function of 
increasing radius (open circles).  The error bar on the most metal-poor data point, 
from the m8 field, has been omitted as it is based on only one stars and therefore 
is not reliable.  The mean metallicity in the three coarser bins representing 
the inner, crossover, and outer regions of M31 are shown as larger filled circles.  
Spectroscopic metallicity measurements for these three components are shown as filled 
squares (the most metal-poor point strongly overlaps the photometric metallicity point).  
The data clearly shows that the halo of M31 is more metal-poor than the inner regions.}
\label{fig:lradius}
\end{figure}

In Figure \ref{fig:lradius} we present the radial metallicity distribution for M31.  The 
mean photometric metallicity in each of our fields is plotted as a function of radius (open circles).  
The mean metallicity of the combined data for the inner, crossover, and outer regions is 
also plotted as larger filled circles.  The data clearly show a trend of decreasing 
metallicity as the distance from M31 increases.  For comparison, we also show the mean 
spectroscopic metallicity in each of our inner, crossover, and outer samples as filled 
squares.  These measurements confirm that M31's outer halo is dominated by stars much more 
metal-poor than the bulge.  Table 3 summarizes our final MDF results.


\begin{table*}
\begin{center}
\caption{}
\begin{tabular}{lcccr}
\hline
\hline
\multicolumn{1}{c}{Field} & \multicolumn{1}{c}{$\langle{R}\rangle$\tablenotemark{1}} & 
\multicolumn{1}{c}{No. M31 RGB}  & 
\multicolumn{1}{c}{$\rm\langle[Fe/H]_{\rm phot}\rangle$\tablenotemark{2}} & 
\multicolumn{1}{c}{$\sigma$} \\ 
& \multicolumn{1}{c}{(kpc)} & \multicolumn{1}{c}{Spheroid Stars} & & \\
\hline
Bulge       & 14 (12--21)  & 127 & $\rm-$0.47$\pm$0.03 ($\rm-$0.57$\pm$0.04)   & 0.39 (0.43) \\
Crossover   & 31 (30--33)  & 87  & $\rm-$0.94$\pm$0.06 ($\rm-$1.12$\pm$0.07)   & 0.60 (0.68) \\
Halo        & 81 (60--165) & 47  & $\rm-$1.26$\pm$0.10 ($\rm-$1.48$\pm$0.11)   & 0.72 (0.73) \\
\hline
\end{tabular}
\tablenotetext{1}{Weighted average of the projected radial distance based on the 
numbers of stars in each individual field that were grouped together, and the range 
of those individual distances (in parentheses).} 
\tablenotetext{2}{Photometric metallicities and dispersions calculated assuming 
[$\alpha$/Fe] = 0.0. Values in parenthesis assume [$\alpha$/Fe] = $+$0.3} 
\label{table:final3}
\end{center}
\end{table*}


\section{Discussion} \label{discussion}

\subsection{Systematic Errors and Measurement Biases} \label{biases}

As discussed in \S\,\ref{intro}, the detection of a radial metallicity
gradient in M31 can directly constrain the global nature of chemical
enrichment during the galaxy formation process.  Our finding, however,
is in contrast to previous observational work on M31 which has found
no significant radial [Fe/H] gradient within $R<30$~kpc (e.g., Durrell
et~al.\ 2001, 2004).  Even the recent \cite{irwin} photometric survey
found no significant color gradient for M31 RGB candidates out to
$R\sim55$~kpc, but uncertainties in statistical background subtraction
make it difficult to interpret the significance of this null result.
Given the relevance of a metallicity gradient and the apparent
discrepancy with earlier studies, we now address and quantify some 
potential biases in our study.  Understanding these effects is crucial
to determining whether or not a spurious metallicity gradient could
have been produced by our target selection and/or analysis procedure.

The target selection for the two innermost fields, H11 and H13s, is
different from the outer fields in the sense that $DDO51$ filter
pre-selection was not used.  Although any selection effect from this
may alter the overall shape of the [Fe/H] distribution, there should
not be any {\it relative\/} bias among different $DDO51$-selected
fields or, for that matter, among non-$DDO51$-selected fields.  To
investigate further whether a global bias exists between the inner and
outer fields we can take advantage of the fact that two of our fields
target the giant southern stream of M31.  The target selection for one
of these fields, H13s, does not involve $DDO51$ photometry, whereas for
the other field, a3, it does.  It is reassuring that both the
\cite{kalirai} study of H13s and the \cite{guh05b} study of a3 found a
substantial population of high metallicity stars in the giant southern
stream despite the $I<22.5$ magnitude cut used for spectroscopic target
selection in both cases (see details below).  A detailed comparison
between these two fields can reveal whether $DDO51$ preselection biases
the sample against the most metal-rich stars.

\begin{figure} 
\begin{center} 
\leavevmode 
\includegraphics[width=8cm]{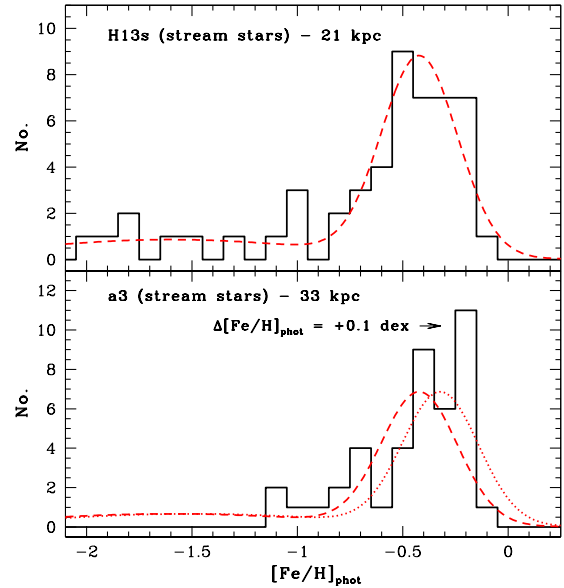} 
\end{center} 
\caption{MDFs for giant southern stream stars in H13s ({\it top\/}) and a3
({\it bottom\/}).  The two MDFs are found to be very similar.  We find that 
the best fit double-Gaussian model to H13s (dashed curve) needs to be shifted only 
by $\Delta$[Fe/H]$_{\rm phot}$ = $\rm+$0.1 dex to fit the 
distribution of stars in the a3 field.  This offset may result from a bias 
related to the way objects were selected in the inner and outer fields in 
our study (see \S\,\ref{biases} for more information).}
\label{fig:mdfstream}
\end{figure}

In Figure~\ref{fig:mdfstream} we present the MDFs in H13s and a3,
including {\it only\/} bona fide stream members.  These have been
selected by imposing velocity cuts on the data.  In H13s,
\cite{kalirai} show that the stream consists of stars moving with
$v_{\rm rad} \leq$ $\rm-$460~km~s$^{-1}$.  Similarly, \cite{guh05b}
show that at the position of a3 stream stars have $v_{\rm rad} \leq$
$\rm-$425~km~s$^{-1}$.  There is some evidence for secondary velocity
peaks in both of these data sets that may or may not be related to the
stream, but we ignore these components here (see Kalirai et al.
2006 for further information).  The resulting two MDFs are remarkably
similar and show a strong peak of metal-rich stars with a small tail
to metal-poor stars.  For H13s, this MDF is well represented by a double
Gaussian (shown as a dashed curve in the upper panel of 
Figure~\ref{fig:mdfstream}).  We now superimpose this same model on
the a3 data (dashed curve in lower panel), and find that the model that 
fits H13s needs to be adjusted only slightly by $\Delta$[Fe/H]$_{\rm phot}$ 
= $\rm+$0.1 dex to achieve the best fit of the stream MDF in a3: this was 
determined by fixing the relative spacing, standard deviations, and relative 
fraction of stars in the two components of the double Gaussian and only
fitting for an overall offset.  This shifted double Gaussian is shown as
a dotted curve in the lower panel.  The comparison of these two
(differently selected) stream samples shows that any systematic [Fe/H]
offset between them is very small.  Since there are several other sources of
systematic errors in [Fe/H] at this level or greater, we ignore this 0.1~dex
offset; in fact, this apparent [Fe/H] offset may well be a result of the
fact that our a3 sample probes deeper down the RGB luminosity function than
the H13s sample (see discussion below).  We note that if we did apply this 
potential offset between $DDO51$-selected and non-$DDO51$-selected data, 
the radial metallicity gradient in M31 would become {\it stronger\/}.

We note that an intrinsic metallicity gradient in the giant southern 
stream is unlikely over such a small radial extent ($R=21$--33~kpc). 
Several investigations have shown that the stream is young and was 
produced less than an orbital period ago (e.g., $\lesssim2$~Gyr---Ibata 
et~al.\ 2004).  The radial extent above represents a very small difference 
in dynamical age and it would be difficult for a metallicity gradient to 
have developed within this timeframe.

Several additional biases exist in our data that would tend to {\it
weaken\/} an observed gradient compared to any true radially-outward 
decrease in the mean metallicity of M31:

\begin{itemize} \item The tip of the RGB is fainter for more metal-rich 
populations (see Figure~\ref{fig:CMD}).  Thus, our magnitude cut for
spectroscopic target selection ($20<I_0<22.5$) can bias the sample
against the inclusion of the most metal-rich RGB stars (Reitzel \&
Guhathakurta 2002).  This bias, if present, is expected to be stronger
for the bulge than for the halo.  This is because our
target selection procedure (see Guhathakurta et al. 2006b) includes
somewhat fainter targets, on average, in the outer fields due to the
sparseness of luminous M31 RGB stars in these regions.

\item The same PDF (see \S\,\ref{cleansample}) is being used to separate 
M31 RGB stars from foreground Milky Way dwarfs in all fields.  This 
selection would tend to pick out RGB stars from the same region of the 
[Fe/H]$_{\rm phot}$ (or location in the CMD or $V{\rm-}I$ color) space 
at all radii (see Gilbert et al. 2006 for details).  Given that our RGB
PDF is based on ``training set'' stars drawn mostly from 
the inner fields, these PDFs favor higher [Fe/H]$_{\rm phot}$ values.  In 
fact, Figure~\ref{fig:pdfs} (d) clearly shows that the RGB training set PDF 
does not sample metal-poor stars very well.  Therefore, the true radial 
trends in metallicity may be even larger than observed.

\item We have assumed a constant age of 12~Gyr for all stars in M31. 
However, the \cite{brown} study indicates that the bulge 
contains a substantial fraction of intermediate age stars.  By contrast,
one expects the halo population to be old by analogy with the Milky
Way's halo.  If in fact the inner region of M31 probed in this study is
younger on average than the halo, then our photometric metallicities
have been underestimated (i.e., assigned to be more metal-poor than they
truly are) in this region. In fact, the comparison between 
[Fe/H]$_{\rm phot}$ and [Fe/H]$_{\rm spec}$ in Figure~\ref{fig:lradius} 
suggests this.  Therefore the adoption of younger isochrones in the 
bulge would lead to higher metallicities for these stars and a larger 
metallicity gradient.  A shift in age from 12 to 6~Gyr translates to a 
$+$0.3~dex offset in [Fe/H]$_{\rm phot}$.

\item We have assumed that all stars in M31 are not enhanced in 
their $\alpha$-element abundances, ([$\alpha$/Fe]).  Stars in the Milky 
Way's halo are known to be $\alpha$-enhanced ([$\alpha$/Fe] = $+$0.3) 
and it is generally believed that this is a result of the stars in the 
halos of galaxies forming in early ``bursts''.  If the M31 halo is also 
$\alpha$-enhanced, then our use of [$\alpha$/Fe] = 0 isochrones has 
overestimated the metallicity for stars in this component (i.e., assigned to 
be more metal-rich than they truly are).  We determined earlier that the 
mean metallicity of our M31 RGB halo sample is in fact $\sim$0.2~dex more 
metal-poor if we assume [$\alpha$/Fe] = $+$0.3 (i.e., we calculate 
$\rm\langle[Fe/H]_{\rm phot}\rangle=-1.5$ for M31's halo under this 
assumption).  

\end{itemize}

\bigskip

These additional biases strengthen our detection of a metallicity gradient 
in M31 are suggest that, if anything, the detected gradient is likely 
{\it smaller\/} than the true gradient. 

\subsection{Bulge/Halo Ratio}\label{bulgehaloratio}

The analysis presented in this paper suggests that, over a large range of
radial distances (12--165 kpc), there is a metallicity gradient within M31. 
This brings up the question: Is this metallicity gradient intrinsic to the
bulge and/or halo, or are we seeing two overlapping M31 structural components 
that are each homogenous but distinct from each other in terms of their chemical
abundance properties?  One way to test these two scenarios is to consider
the crossover region at 30~kpc, where we find a mean metallicity of 
$\rm\langle[Fe/H]_{\rm phot}\rangle=-0.94\pm0.06$.  \cite{guh06a} showed that at
this radius there is roughly an equal mix of bulge and halo stars based
entirely on fits to the minor-axis surface brightness profile of M31.  If
our data are consistent with the latter scenario (e.g., a homogenous bulge
and halo), then the 30~kpc MDF should in fact be an equal superposition of
the distinct bulge and halo MDFs.  To test this, we randomly selected 47 bulge 
stars and grouped them together with the same number of stars in the halo 
sample.  The resulting composite MDF is compared to the 30~kpc MDF derived 
from the a0 and a3 fields and found to be very similar.  A two-sided 
K-S test (see \S\,\ref{kstest}) between this composite ``bulge+halo'' MDF 
and the 30~kpc MDF returns a probability of the null hypothesis being 
true of $P >$  0.50 (i.e., of the two samples being drawn from the same 
parent population).

We therefore conclude that our data are consistent with a chemically
homogenous bulge and halo whose relative contributions change
systematically with radius and give rise to the observed metallicity
gradient.  This is not to say that M31's bulge and halo are perfectly
homogenous in a large-scale radial sense---only that our present samples
lack the size and precision to detect any subtle metallicity gradients that
may be intrinsic to the bulge or halo.  For example, Figure~\ref{fig:lradius} 
shows a very mild trend that an inverted metallicity gradient may be 
present in the halo of M31.  However, splitting our 60--165~kpc 
halo fields into two bins (e.g., $R <$ 100~kpc and $R <$ 100~kpc) 
suggests that such a trend is statistically not significant (marginally).  
This measurement is difficult to make given the small number of stars.  
Future samples will no doubt be able to rectify this situation and allow 
us to characterize radial trends in the intrinsic chemical abundance 
properties of the bulge and halo in detail.

\subsection{Relation of our Fields to M31's Disk}

None of the fields used in this study are contaminated by M31's extended 
disk to any significant degree.  \cite{ibata05} find that this disk is a 
low-surface brightess, kinematically cold (velocity dispersion, 
$\sim$30~km~s$^{-1}$) structure at $R$ = 15--40~kpc and is an extention of 
the inner disk.

Our fields tend to lie close to M31's minor axis, and are therefore well 
removed from the major axis.  The most susceptible fields in our study 
that could potentially suffer from this contamination are the two innermost 
pointings, H11 and H13s.  The projected distance of H13s is 21~kpc, which in 
the plane of the disk is $>$80~kpc.  Therefore this field is located well 
beyond the extendend disk and \cite{kalirai} have already ruled out 
any significant contribution to this field by a smooth disk population.  We 
now focus on H11, which is located at $R$ = 12~kpc in the bulge of M31 
(i.e., 50~kpc in the disk).  We can rule out disk star contamination in 
this field based on several independent arguments,

\begin{itemize} 

\item \cite{guh06a} present the surface brightness profile of M31 and 
show that at the position of H11, disk stars are expected to be greatly 
outnumbered by bulge stars.  The bulge contribution to H11 is at least 
an order of magnitude greateer than the disk contribution.

\item Figure~\ref{fig:kinematics} (Rich et~al. 2006, in preparation) shows 
that the velocity histogram of H11 is well represented by a broad Gaussian 
centered near M31's systemic velocity.  There is no evidence for a 
30~km~s$^{-1}$ kinematically cold population (as seen in the fields that 
probe the extended disk of M31 in Ibata et~al. 2005).  In fact, the 
velocity dispersion of M31 RGB stars in H11 is measured to be 
approximately three times higher than that of M31's extended disk.  A 
maximum likelihood analysis of the H11 radial velocity data indicates 
that it is impossible to hide a substantial fraction of the stars in a 
kinematically cold (30~km~s$^{-1}$) component (Reitzel et~al. 2006, in 
preparation).  Although this extended disk population dominates Ibata et 
al.'s RGB samples in fields near the major axis, their two minor axis 
fields, F05 and F07, also show no evidence of this population, as noted 
by the authors.

\item Another scaling argument based on kinematics can be used to rule out 
the presence of a substantial fraction of M31 disk members in the 
H11 field.  Figure~9 in \cite{kalirai} shows radial velocity histograms 
for H11 ($R$ = 12~kpc, S-E minor-axis), H13s ($R$ = 21~kpc, giant 
southern stream), and H13d ($R$ = 25~kpc, N-E major-axis).  Consistent 
with \cite{ibata05}, our H13d field shows a kinematically cold population 
of stars, presumably M31's disk.  This field lies at 25~kpc in the disk 
plane, and also at 25~kpc in the bulge (as it is on the major axis). We 
find that two-thirds of the RGB stars in this field belong to the cold 
component (disk) and one-third are part of a dynamically-hot (bulge) 
component (Reitzel et al. 2006, in preparation).  By comparison, our 
H11 field is located at 50~kpc in the disk plane and 20~kpc in the 
bulge (accounting for the 5:3 flattening of the bulge).  Thus, the H11 
field is twice as far out in the disk as H13d (ten versus five disk scale-lengths) 
and at a slightly smaller radial distance in the bulge.  Moving out in the 
disk by five exponential scale lengths reduces the disk contribution by 
over two orders of magnitude.  Thus, very few, if any, disk members should be 
present in our H11 field.

\item T. Brown et al. (2006, in prep) have recently found evidence 
for {\it younger\/} stars in M31's disk that are {\it not\/} seen in 
H11.  By obtaining ultra-deep {\it HST\/}/ACS observations of stars in 
both of these fields (directly overlapping our pointings), Brown et al. 
are able to reach the main-sequence turnoff in M31's bulge and disk.  
The reconstructed star formation histories of the these two components 
indicate some clear differences.  For example, the disk field contains 
a younger main sequence and is generally more metal-rich than the 
bulge field.  Therefore, these independent observations also suggest 
that our H11 field is not contaminated with disk stars and rather 
represents M31's bulge.

\end{itemize}

Therefore, the surface brightness, kinematics, ages, and metallicities of stars 
in H11 argue against them being part of a disk like component in M31.  We 
note that our next closest field, a0, is located at $R sim$ 30~kpc in M31's 
halo (i.e., over 130~kpc in the disk) and therefore does not sample the 
extended disk.

\cite{worthey} make the radical hypothesis that a thick
disk dominates the entire region within $R <$ 50~kpc in M31.  This
hypothesis was put forward to explain the high mean metallicity seen
in previous studies of M31, but was not based on any kinematical data.
Reitzel et al. (2006, in preparation) show that the stellar kinematics
in the inner regions of M31 are inconsistent with Worthey et al.'s
hypothesis, and confirm that the H11 and H13s samples used in this
work represent M31's bulge population instead.


\subsection{Metallicity Distribution of the M31 Bulge: Comparisons to Other
Studies} \label{mdfcompothers}

Although the halo sample of M31 RGB stars presented in this work is 
unique, a limited number of studies have probed the MDF of the bulge 
of M31 (although these studies have often referred to this component as the 
``halo'').  The best study comes from ultra-deep {\it HST\/}/ACS observations 
by \cite{brown2} who have studied both the bulge and giant southern stream.  
These data reach faint enough to detect the lower RGB, 
horizontal branch, and main-sequence turnoff of M31.  At these fainter 
magnitudes, there is minimal foreground dwarf star contamination.  The 
{\it HST\/} CMDs of these two fields show striking similarities over all of 
the above phases of stellar evolution suggesting that the ages and 
metallicities of stars in the two fields are very similar.  Furthermore, 
our H11 and H13s Keck/DEIMOS spectroscopic pointings have been carefully 
chosen to directly overlap these two {\it HST\/}/ACS fields. In \cite{brown}, 
it was shown that the H11 CMD contains both metal-poor and metal-rich stars 
($\rm[Fe/H]>-0.5$). Given the striking similarity between the {\it HST\/} 
CMDs in these two fields, the H13s field must have a comparable MDF.  Our 
measurement of the MDF in H11 and H13s confirm this (see 
Figure~\ref{fig:cummdfphot}), both showing a distribution of stars 
near $\rm[Fe/H] \sim -0.5$.

The MDF of M31's bulge has also been probed photometrically by 
\cite{durrell01} ($R\sim20$~kpc) and \cite{durrell04} ($R\sim30$~kpc). 
These measurements are based on wide field CFHT data.  Durrell et al. 
compute the MDF by interpolating the location of stars on the CMD within a 
grid of $\alpha$-enhanced RGB models.  This is done both for the target 
fields and for well-removed control fields.  The final MDF is obtained by 
subtracting the control field MDF from the target field MDF.  The results,
in both fields, indicate that M31's bulge is dominated by relatively 
high metallicity stars, $\rm\langle[m/H]\rangle\sim-0.5$
($\rm\langle[Fe/H]\rangle=-0.8$, for [$\alpha$/Fe] = $\rm+$0.3).  As Table~3 
shows, our inner field is located interior to Durrell et al.'s 20~kpc 
pointing, and our crossover field is located slightly exterior to Durrell et al.'s 
30~kpc pointing.  Our measured metallicities, assuming the same [$\alpha$/Fe] = 
$\rm+$0.3 adopted by Durrell et al., are found to nicely bracket Durrell et al's 
metallicity (see Table~3).  This suggests that Durrell et al. have succeeded 
in statistically eliminating Milky Way dwarf stars with their use of an 
appropriate control field.  Of course, this becomes more difficult as the 
distance from M31 increases, and therefore the methods discussed in 
\cite{gilbert} become more important.

Other studies of M31 (e.g., Worthey et~al. 2005, Bellazzini et~al. 2003) 
have similarly concluded that the bulge is dominated by metal-rich stars
($\rm\langle[Fe/H]\rangle\sim-0.5$).  Both these studies, and those 
discussed above,  vary from one another in terms of both the source of 
the data and the types of models used to determine metallicity.  
Therefore, external comparisons of these results to our MDFs for H11 and 
H13s may not yield a perfect agreement and are not as powerful as relative 
comparisons of the MDF at various radii from a single data set (as presented 
in this paper).  However, it is encouraging that such an external comparison 
does in fact yield similar metallicities and thus suggests that net effect 
of systematic errors in our MDFs is small.

\subsection{The M31 Halo Versus the Milky Way Halo} \label{m31vsgalaxy}

In \S\,\ref{intro} we discussed the assembly of massive galaxies such as the
Milky Way and M31 in the context of hierarchical merging of smaller galaxies. 
Several predictions arise from this theoretical framework that have, until
now, been challenged by our understanding of M31.  Both structurally and 
in terms of chemical enrichment, the M31 ``halo'' was believed to be
different from our own Galactic halo and different from the predictions of
generally-accepted halo formation models.  The wide-field imaging survey of
\cite{ostheimer} was the first to find preliminary evidence of a metal-poor
stellar halo in M31: it used $DDO51$ selection to greatly reduce foreground
contamination by Milky Way dwarf stars and was thereby able to probe out to
larger radii than previous studies.  By coupling the Ostheimer $DDO51$
photometry with Keck/DEIMOS spectroscopy, \cite{guh06a} showed that M31 does
possess a stellar component that is structurally distinct from its bulge and
whose radial extent ($R>160$~kpc) and surface brightness profile ($\sim
R^{-2}$) resemble those of our Galaxy's stellar halo.

In this paper we have taken another step toward bridging the apparent
disparity between M31 and the Galaxy/galaxy formation models by showing
that this newly-discovered stellar halo in M31 is chemically distinct from
its bulge, and is in fact quite metal poor: 
$\rm\langle[Fe/H]_{\rm phot}\rangle=-1.5\pm0.1$ ($\sigma$ = 0.7).  For 
comparison, \cite{morrison} have recently measured metallicities from spectra 
of Milky Way halo RGB stars located at distances between 15 and 83~kpc from 
the Galactic center.  The mean metallicity of this sample is measured to 
be $\rm\langle[Fe/H]\rangle=-1.6$ ($\sigma$ = 0.6), in good agreement with 
earlier studies.  Therefore we find that the mean metallicity, and spread, 
of the M31 halo is similar to that of the Galactic halo.

\bigskip

The results presented in this paper show that a survey out to 20--30~kpc 
is not sufficient to properly isolate the M31 halo.  In the 
Milky Way, the halo begins to dominate the thin disk at a radius interior to 10~kpc, 
locally (Majewski 1993).  Therefore, our sampling distance in M31 is not 
comparable to the Milky Way.  M31 likely contains a substantially larger 
bulge than the Milky Way and this bulge dominates over the halo in all 
previous studies of M31 that were limited to the central few degrees 
of the galaxy.  Our data show clear
evidence of a halo and bulge in M31 that are distinct from each other in
terms of both structure and chemical abundance properties.

Prior to the detection of 47 M31 halo stars in this analysis, very few 
(if any) spectroscopically confirmed RGB stars had been detected in 
M31 with $R \gtrsim$ 60~kpc.  Future observations will allow us to 
gather larger bulge and halo samples to establish 
whether or not these components contain {\it intrinsic\/} radial 
metallicity gradients.  A detailed study of the {\it true halo\/} of M31 will 
require new observational strategies to gain significant insights into 
this important, but tenuous/elusive, stellar population.

\section{Conclusions} \label{conclusion}

Imaging and multiobject spectroscopic surveys of M31 have recently brought to 
light several unanticipated results.  By isolating M31 RGB stars from Milky Way 
foreground dwarf contamination using a new technique, we are able to probe further 
into the outskirts of M31 than previous studies ($R$ = 12--165~kpc).  In a 
separate paper we established the existence of a true stellar halo in M31, 
thereby resolving a long-standing concern that the surface brightness 
profile of M31 looks different from the Milky Way.  
In this paper, we analyze the metallicity distribution function 
of stars in twelve fields spanning this large radial extent.  Our data show
a radial metallicity gradient in the spheroid (bulge and halo) of M31.  
The outer region of M31 ($R >$ 60 kpc), dominated by its halo, is composed 
of stars deficient in metals relative to those in the inner region.  Based on 
the sample of 47 stars, we find a mean chemical abundance of
$\rm\langle[Fe/H]_{\rm phot}=-1.48\pm0.11$ ($\sigma$ = 0.73) for the newly 
discovered halo of M31 (assuming [$\alpha$/Fe] = $+$0.3).  This confirms 
predictions from current galaxy formation models which suggest that the 
inner regions of large galaxies should be chemically enriched relative 
to the outer regions.


\acknowledgments

We wish to thank Peter Stetson and Jim Hesser for help in acquiring the
CFHT/MegaCam imaging fields for this project.  We are also grateful to Peter
Stetson and James Clem for providing programs and for many useful discussions
regarding the astrometry of the CFHT images.  We wish to thank Carynn Luine
for help with the verification of the radial velocity measurements and Leo
Girardi for providing us with an extensive grid of theoretical stellar
isochrones.  We also acknowledge Tom Brown for making available unpublished 
results based on Cycle 13 {\it HST\/} observations.  J.S.K. is supported 
by NASA through Hubble Fellowship grant HF-01185.01-A, awarded by the Space
Telescope Science Institute, which is operated by the Association of 
Universities for Research in Astronomy, Incorporated, under NASA 
contract NAS5-26555.  This project was also supported 
by NSF grant AST-0307966 and NASA/STScI grant GO-10265.02 (J.S.K., P.G., and 
K.M.G.), an NSF Graduate Fellowship (K.M.G.), NSF grant AST-0307931 (R.M.R. and 
D.B.R.), and NSF grants AST-0307842 and AST-0307851, NASA/JPL contract 1228235, 
the David and Lucile Packard Foundation, and The F.~H.~Levinson Fund of the 
Peninsula Community Foundation (S.R.M.).

\end{document}